\documentclass[aps,prb,reprint,superscriptaddress]{revtex4-1}
\usepackage{graphicx}
\usepackage{amsmath,amsfonts}
\usepackage{xcolor}
\definecolor{Red}{rgb}{0.9,0.0,0.1}
\definecolor{Blue}{rgb}{0.1,0.1,0.9}
\usepackage{natbib}

\makeatletter
\newcommand*{\citenst}[2][]{%
  \begingroup
  \let\NAT@mbox=\mbox
  \let\@cite\NAT@citenum
  \let\NAT@space\NAT@spacechar
  \let\NAT@super@kern\relax
  \renewcommand\NAT@open{[}%
  \renewcommand\NAT@close{]}%
  \cite[#1]{#2}%
  \endgroup
}
\makeatother

\begin{document}

%\preprint{}

\title{Interaction between magnetic vortex cores in a pair of nonidentical nanodisks}


%\author{}

\affiliation{Centro Brasileiro de Pesquisas F\'{\i}sicas, 22290-180, Rio de Janeiro, RJ, Brazil }

\author{J.P. Sinnecker}

\author{H. Vigo-Cotrina}

\author{F. Garcia}

\author{E.R.P. Novais}

\author{A.P. Guimar\~aes}
%\affiliation{Centro Brasileiro de Pesquisas F\'{\i}sicas, 22290-180, Rio de Janeiro, RJ, Brazil }
\email[Author to whom correspondence should be addressed:]{apguima@cbpf.br}

\date{\today}

\begin{abstract} Interacting magnetic nanoobjects constitute one key component for many proposed spintronic devices, from microwave nano-oscillators to magnetic memory elements. For this reason, the mechanism of this interaction and its dependence with distance $d$ between these nanoobjects has been the subject of several recent studies. In the present work, the problem of the interaction between magnetic nanodisks with magnetic vortex structures is treated both analytically and through micromagnetic simulation. The coupling of two nonidentical magnetic nanodisks, i.e., with different gyrotropic frequencies, is studied. From the analytical approach the interactions between the nanodisks along $x$ and $y$ directions (the coupling integrals) were obtained as a function of distance. From the numerical solution of Thiele's equation we derived the eigenfrequencies of the vortex cores as a function of distance. The motion of the two vortex cores, and consequently the time dependence of the total magnetization $M(t)$ were derived both using Thiele's equation and by micromagnetic simulation. From $M(t)$, a recently developed method, the magnetic vortex echoes, analogous to the NMR spin echoes, was used to compute the distance dependence of the magnetic coupling strength. The results of the two approaches differ by approximately $10\%$; using one single term, a dependence with distance found is broadly in agreement with studies employing other techniques.
\end{abstract}

\maketitle

\section{Introduction}

Magnetic objects of nano- and mesoscopic dimensions of different shapes - squares, ellipses or disks - may have, as their ground state, a vortex structure \cite{Guimaraes:2009, Chien:2007, Guslienko:2008b, Novais:2011, Soares:2008}. This state is characterized by magnetization in the plane of the nanostructure, tangential to concentric circles, and a small core where the magnetization is perpendicular to the plane. One can define the circulation $c=+1$ for counterclockwise (CCW) in-plane magnetization direction, or $c=-1$ for CW direction; the polarity is $p=+1$ for magnetization of the vortex core along the $+z$ axis, and $p=-1$ for the opposite direction ($-z$). The physical description of the vortex properties is usually made within two analytical models: rigid vortex model \cite{Usov:1994} and the two-vortex model \cite{Guslienko:2001,Metlov:2002}.

Magnetic structures with vortices have many potential applications, e.g., as spin-torque nano-oscillators (STNO's) \cite{Lehndorff:2009}, magnetic random memories (MRAM's) \cite{Bohlens:2008, Pigeau:2010}, or logic gates \cite{Jung:2012}. The applications usually require magnetic elements arranged in a regular array where the characterization of the interaction between them
is required: in some cases it is necessary for the functioning of the device, in other cases it has to be minimized. This interaction allows the coupling of the nanoobjects\cite{Shibata:2003} and the loss-less transmission of energy\cite{Jung:2011}.

When a magnetic vortex structure, for example, a magnetic nanodisk, is in equilibrium, its vortex core rests at its center, and the structure has magnetic flux-closure. In this configuration the coupling with nearby nanoelements is minimum. Conversely, when the vortex structure is out of equilibrium, with its core displaced e.g., by an external magnetic field or a spin-polarized current, magnetostatic coupling with the neighbor elements results. The dependence with distance of this coupling has been the subject of several studies in recent years \cite{Sugimoto:2011, Jung:2011, Sukhostavets:2011, Garcia:2012}.

Once the excitation of the vortex cores through an external agent is over, they return to their equilibrium position, performing a periodic spiral-like trajectory. This motion, called gyrotropic motion, has been described analytically through Thiele's equation, that is derived from Landau-Lifshitz equation \cite{Huber:1982}, and has also been obtained from micromagnetic simulations.
The angular frequency of this motion depends on the saturation magnetization of the material and on the aspect ratio of the disks.
It is typically in the range of hundreds of MHz.
Also, several experiments using different techniques have expanded our knowledge of this phenomenon, e.g., \citenst{Park:2003}.

\begin{figure}[!ht]
\centering
\includegraphics[width=1\columnwidth]{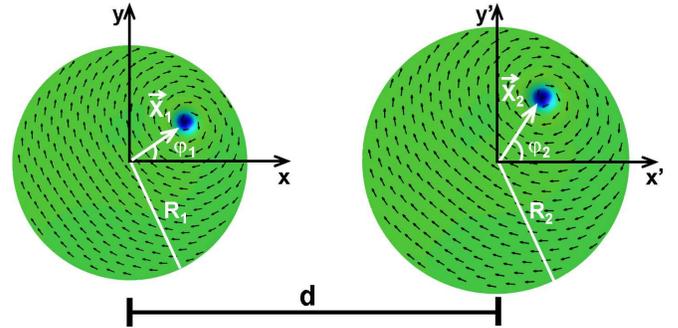}
\caption{Coupled disks with vortex magnetic configuration and different radii $R_1$ and $R_2$, separated by a center to center distance $d$.}\label{fig:CoupledDisks}
\end{figure}

All these considerations justify why vortex core dynamics, and vortex core interactions, have recently attracted the interest of researchers in the area of Nanomagnetism. Recent studies, both theoretical and experimental, explore the vortex core interaction and vortex dynamics in identical disks pairs\cite{Jung:2011,Jung:2010,Shibata:2003,Sugimoto:2011,Sukhostavets:2011}.

The simplest system where one can study interacting nanodisks is, of course, a pair of such magnetic structures; it is
therefore the ideal system for the investigation of the properties of the interaction, its dependence with distance, etc.

Some devices were proposed using nanodisks with different diameters, e.g., in magnonic devices\cite{Huber:2011} or in nano-oscillators\cite{Belanovsky:2013}, although there are few studies of the interactions in more complex structures, such as nonidentical disk arrays in which each element interacts with all the others\citep{Garcia:2012}.

In the present work, the problem of the interaction between pairs of magnetic
nanodisks with magnetic vortex structures and different gyrotropic frequencies, is analyzed both analytically and through micromagnetic simulation. The present discussion is applicable to pairs of magnetic nanodisks
that have different gyrotropic frequencies, arising either from different radii, or different materials, or different thicknesses. This may be
relevant to the study of fabricated pairs of magnetic nanodisks, where a distribution of frequencies is inherent in the actual samples.
We will choose as illustration the
difference in radii, as shown in Fig. \ref{fig:CoupledDisks}.

Here we generalized the analytical treatment of the formulation of disk interaction, for any pair of disks.
Our results can be well described by interaction intensities that are a multipole expansion with terms of the form $d^{-n}$, with $n=$3, 5, 7 and 9, i.e., dipole-dipole, dipole-octupole, octupole-octupole and dipole-triacontadipole interactions, respectively, as recently demonstrated by Sukhostavets \textit{ et al.} \cite{Sukhostavets:2013}.

Finally, we used the recently reported magnetic vortex echo method \cite{Garcia:2012}, in order to obtain information on the interaction between the disks, using the magnetization
$M(t)$ given by two different approaches, on the one hand using Thiele's equation, and on another using micromagnetic simulation.

\section{Results and discussion}
\label{section:analytical}

\subsection{Analytical description}

%- generalization for $R1\neq R2$
The analytical description of the interaction between two disks starts by considering both of them with vortex magnetic configuration, and with diameters $2R_1$ and $2R_2$, thickness $L$ and with centers separated by a distance $d$ along the $x$ axis, as shown in Fig. \ref{fig:CoupledDisks}.
The magnetostatic interaction between the disks is due to the occurrence of magnetic charges $\sigma_i (i = 1,2)$ on their surfaces (top, bottom and lateral) induced by the shift of the vortex cores from the equilibrium positions \cite{Guslienko:2001}. These charges are given in the rigid vortex model by\cite{Guslienko:2001,Shibata:2003}:

\begin{eqnarray} \sigma_i = -\frac{c_iM_s(x_i\sin{\varphi_i} -y_i\cos{\varphi_i})}{\sqrt{R_i^2 + |\textbf{X}_i|^2 - 2R_i(x_i\cos{\varphi_i} + y_i\sin{\varphi_i})}}\label{eq:sigmai} \end{eqnarray}

\noindent where $c_i = \pm 1$ is the $i^{th}$ disk circulation, $M_s$ is saturation magnetization and $\textbf{X}_i$, $\phi_i$ and $R_i$ are defined according to Fig. \ref{fig:CoupledDisks}. The magnetostatic interaction energy $W_{int}(\textbf{X}_1,\textbf{X}_2)$ of the side surfaces of two disks is \cite{Shibata:2003, Sukhostavets:2011}:

\begin{eqnarray}
 W_{int}(\textbf{X}_1,\textbf{X}_2) = \frac{1}{2}\int dS_1\int dS_2\frac{\sigma_1 \sigma_2}{|\textbf{r}_1 - \textbf{r}_2|}\label{eq:Wi}
\end{eqnarray}

The integration is performed over the surfaces $S_1$ and $S_2$ of each disk\cite{Sukhostavets:2011}, $dS_i = R_idz_id\varphi_i$, and $\textbf{r}_1 = \textbf{r}$, $\textbf{r}_2 = \textbf{r}' + d\hat{x}$, as defined by reference \citenst{Sukhostavets:2011}.

Inserting Eq. \ref{eq:sigmai} into Eq. \ref{eq:Wi} and considering $|\textbf{X}_i|/R \approx 0$
(the vortex displacement is much smaller than the disk radius) we have:

\begin{eqnarray}
W_{int}(\textbf{X}_1,\textbf{X}_2) = c_1c_2(\eta_x x_1 x_2 + \eta_y y_1y_2) + \mathcal{O}(|X_i|^3)
\end{eqnarray}

\noindent with:

\begin{eqnarray}
\eta_{x,y} = \frac{\mu_0M_s^2\bar{R}}{8\pi}I_{x,y}
\end{eqnarray}
\noindent where\\
\begin{eqnarray}
I_x = \int T\sin{\varphi_1} \sin{\varphi_2}d\varphi_1 d\varphi_2d\bar{z}_1d\bar{z}_2\label{eq:etax}\\
I_y = \int T\cos{\varphi_1} \cos{\varphi_2}d\varphi_1 d\varphi_2d\bar{z}_1d\bar{z}_2\label{eq:etay}
\end{eqnarray}
\noindent with\\
\begin{eqnarray*}
T & = & \left[ g_1^2 + g_2^2 + \bar{d}^2 \right.\\
    & & - 2g_1g_2\cos{(\varphi_2 - \varphi_1)} \\
    & & + 2\bar{d}(g_2\cos{\varphi_2} - g_1\cos{\varphi}_1) \\
    & & \left. + (\bar{z}_1 + \bar{z}_2)^2\right]^{-1/2}
\end{eqnarray*}
Here we have considered the dimensionless variables: $ g_i=R_i/\bar{R}$,
$ \bar{z}_i=z_i/\bar{R}$, $\bar{d}=d/\bar{R}$ with $\bar{R} = (R_1 + R_2)/2$ for $i = 1,2$.\\

The limits of integration are from 0 to $2\pi$ in $\phi_1$, $\phi_2$, and from 0 to $L/\bar{R}$ in $z_1$, $z_2$.

In Eqs. \ref{eq:etax} and \ref{eq:etay}, $I_x$ and $I_y$ describe the interactions along $x$ and $y$ directions between two disks and can be found by numerical integration. Eqs. \ref{eq:etax} and \ref{eq:etay} are a generalization of similar results obtained previously for a pair of coupled identical disks \cite{Shibata:2003, Sukhostavets:2011}, considering now nonidentical disks.

\begin{figure}[!ht] \includegraphics[width=1\columnwidth]{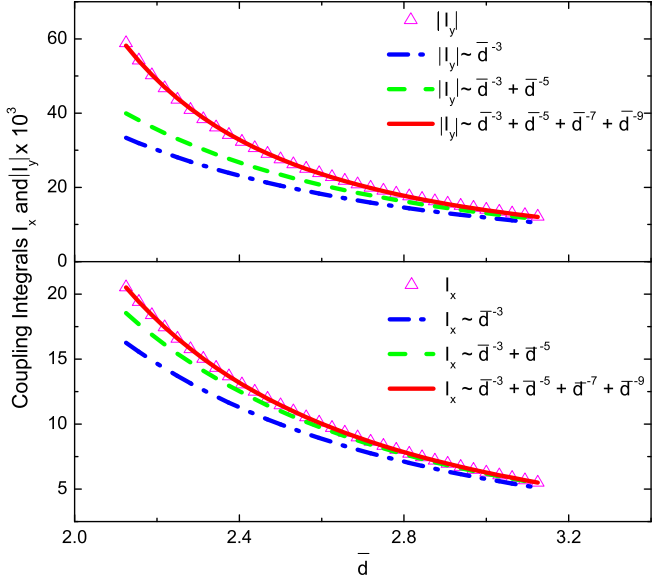} \caption{Coupling integrals $I_x$ and $\mid{I_y\mid}$ as a function of the reduced distance $\bar{d}=d/\bar{R}$ for disks with $L = 20\:$nm and $R_1 = 150\:$nm and $R_2 = 170\:$nm calculated for 340\,nm $< d <$ 500\,nm; the red continuous line is the best fit to a multipole expansion with terms up to $n=9$ ; the green dashed
line represents the result of a multipole expansion, but considering only the dipolar-dipolar and dipolar-octupolar contributions; the blue dash-dot line is the result of a multipole expansion, considering only the dipolar-dipolar term.}\label{fig:EtaxEtayVsD} \end{figure}

Fig. \ref{fig:EtaxEtayVsD} shows $I_x$ and $I_y$ calculated for a separation $d$ between the disks
in the range 340\,nm $< d <$ 500\,nm.

According to Sukhostavets \textit{et al.} \cite{Sukhostavets:2013}, the interaction between a pair of equal disks can be described using interaction coefficients that depend on the center to center disk distance, as a multipole magnetostatic interaction expansion
%of the form $Ad^{-3} + Bd^{-5} + Cd^{-7} + Dd^{-9}$
where the only non-zero terms have odd exponents and the most important interactions to be taken into account for coupled disk dynamics are dipole-dipole, dipole-octupole and octupole-octupole. The coefficients $A,B,C$ and $D$ are different for interactions along the $x$ and $y$ directions (see \citenst{Sukhostavets:2013}, Eq. 16).

In our results, the interactions in the $x$ and $y$ directions are given by the coupling integrals $I_x$ and $I_y$, respectively; the difference between these
integrals arises naturally, from the symmetry of the problem. The interaction is a function of the distance between the disks, as can be seen in Fig. \ref{fig:EtaxEtayVsD}. As the distance increases, $I_x$ and $I_y$ decrease.

The coupling integrals $I_x$ and $I_y$, when fitted to a single term of the form $\bar{d}^{-n}$, lead to values of $n = 3.41 \pm 0.02$ and $n = 4.08 \pm 0.07$, for $I_x$ and $I_y$, respectively. These values of $n$ are similar to those found by Sukhostavets \textit{et al.}\cite{Sukhostavets:2013} and Garcia \textit{et al.}\cite{Garcia:2012} for equivalent parameters.

For a smaller distance $d$, between $340$ and $370\:$nm, we have $n = 3.70 \pm 0.02$ and $n = 5.05 \pm 0.09$ for $I_x$ and $I_y$, respectively. As the disk distance becomes smaller, the interaction increases and higher-order interaction terms such as dipole-octupole, octupole-octupole terms have a higher contribution, increasing the value of $n$, as recently obtained\cite{Sukhostavets:2013} for identical disks.

The dependence of $I_x$ and $I_y$ on $\bar{d}$ in Fig. \ref{fig:EtaxEtayVsD} can also be described with a multipole expansion with odd terms of the reduced distance $\bar{d}$ between the disk centers as:

\begin{equation}
\label{multipole}
I_{x,y} = A\bar{d}^{-3} + B\bar{d}^{-5} + C\bar{d}^{-7} + D\bar{d}^{-9}
\end{equation}

\noindent where the terms of form $\bar{d}^{-n}$, with $n=$3, 5, 7 and 9 are again the dipole-dipole, dipole-octupole, octupole-octupole and dipole-triacontadipole interactions, respectively\cite{Sukhostavets:2013}.

A good fit can be found when considering all terms, as can be observed in the red continuous line of Fig. \ref{fig:EtaxEtayVsD}. No reasonable fit can be found by only considering the dipole-dipole interaction $(A\bar{d}^{-3})$, especially for close disks $(d \leq 350\:$nm). We can estimate the relevance of each term in Eq. \ref{multipole} by plotting the curves obtained using the coefficients from the multipole expansion best fit, but considering only the dipole-dipole interaction term (dash-dot blue line in Fig. \ref{fig:EtaxEtayVsD}) or the dipole-dipole plus dipole-octupole terms (dashed green line in Fig. \ref{fig:EtaxEtayVsD}).

%Table \ref{results} summarizes the best fitting results for disks with equal diameters ($d=XXX$\,nm) and non equal diameters ($d_1=XXx\,nm$ and $d_2=XXx\,nm$). The table also shows the %results obtained by Sukhostavets for a pair of equal disks. A good agreement with Sukhostavets theory is obtained for a pair of identical disks.

\subsection{Numerical solution of Thiele's equation}
\label{subsection:numericalThiele}

One interesting aspect of the coupled nanodisk pair studies is the determination of the vortex gyrotropic eigenfrequencies. These frequencies can be determined analytically using the linearized Thiele's equation, that can be written, considering zero damping \cite{Hubert:1999}:

\begin{eqnarray}
\textbf{G}_i \times \frac{d\textbf{X}_i}{dt} - \frac{\partial W (\textbf{X}_1,\textbf{X}_2)}{\partial \textbf{X}_i} = 0\label{eq:Thiele}
\end{eqnarray}

\noindent where $\textbf{G}_i$ is the gyrovector, $\textbf{G}_i = - G_ip_i\hat{z}$, $G_i = 2\pi \mu_0M_sL_i/\gamma$, $\gamma$ is the gyromagnetic ratio and $M_s$ is the saturation magnetization.

The potential energy is

\begin{eqnarray*}
W (\textbf{X}_1,\textbf{X}_2) & = & W_1(\textbf{X}_1) \\
                                             & &+ W_2(\textbf{X}_2)\\
                                             & &+ W_{int}(\textbf{X}_1,\textbf{X}_2).
\end{eqnarray*}

$W_1(\textbf{X}_1)$ and $W_2(\textbf{X}_2)$ are the potential energies of each isolated disk. $W_i(\textbf{X}_i) = W_i(0) + \kappa_i \textbf{X}^2_i/2$, where $W(0)$ is the potential energy for $\textbf{X}_i = (0,0)$ and $\kappa_i = 40\pi M_s^2L_i^2/9R_i$ is the stiffness coefficient \cite{Guslienko:2001}. $W_{int}(\textbf{X}_1,\textbf{X}_2)$ is the magnetostatic interaction between the disks.

The system of Thiele's equation of motion (Eq. \ref{eq:Thiele}) is simplified using a solution $\textbf{X}_i(t) = \textbf{X}_i(\omega)\exp(i\omega t)$, where $\omega$ is the frequency, thereby obtaining a matrix equation of the form $\hat{B}\textbf{A} = i\omega\textbf{A}$:

\begin{equation}
\begin{bmatrix}
0 & - \omega_1p_1 & 0 & -b\omega_1p_1 \\
\omega_1p_1 & 0 & a\omega_1p_1 & 0 \\
0 & -d\omega_2p_2 & 0 & -\omega_2p_2\\
c\omega_2p_2 & 0 & \omega_2p_2 & 0
\end{bmatrix}
\begin{bmatrix}
x_1\\
y_1\\
x_2\\
y_2
\end{bmatrix}
= i\omega
\begin{bmatrix}
x_1\\
y_1\\
x_2\\
y_2
\end{bmatrix}\label{eq:MatrixBA}
\end{equation}

\noindent where $a = c_1c_2\eta_x/G\omega_1$, $b = c_1c_2\eta_y/G\omega_1$, $c = c_1c_2\eta_x/G\omega_2$ and $d = c_1c_2\eta_y/G\omega_2$ with the eigenfrequency of each isolated disk $\omega_i = \kappa_i/G_i$ (i = 1,2).

From Eq. \ref{eq:MatrixBA} we get the coupling frequencies $\omega_i$:

\begin{eqnarray}
(\omega^p_{+,-})^2 = \frac{{\omega_1}^2 + {\omega_2}^2 + 2bcp\omega_1\omega_2
\pm \sqrt{\bigtriangleup}}{2}\label{eq:omega}
\end{eqnarray}
\noindent where
\begin{eqnarray*}
\bigtriangleup = ({\omega_1}^2 - {\omega_2}^2)^2 + 4{\omega_1}^2{\omega_2}^2 (ca + bd) \\
+ 4p\omega_1\omega_2bc ({\omega_1}^2 + {\omega_2}^2)
\end{eqnarray*}

\noindent with $p =p_1p_2$; note that in this expression the circulations $c_1$ and $c_2$ only appear squared, and therefore
the frequencies do not depend on the sign of the circulations $c_i$.

\begin{figure}[!ht] \centering \includegraphics[width=1\columnwidth]{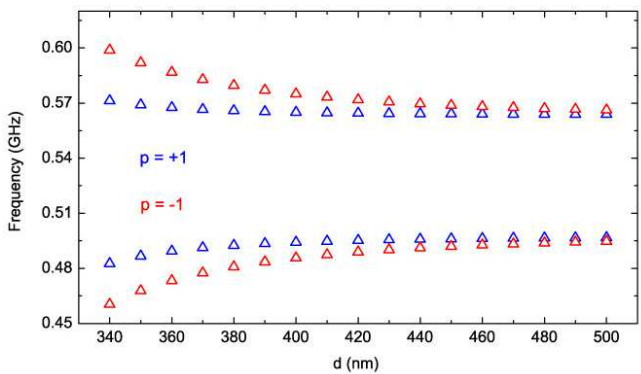} \caption{Variation of coupling frequencies $\omega^p_{\pm}/2\pi$ with the separation distance $d$ between two disks, with radii $R_1 = 150\:$nm and $R_2 = 170\:$nm, thickness $20\:$nm, and combined polarities $p=p_1p_2=+1$ and $-1$. These values were obtained from Eq. \ref{eq:omega}.}\label{fig:CouplingVsD} \end{figure}

We considered a disk pair with radii $R_1 = 150\:$nm, $R_2 = 170\:$nm and thickness $20\:$nm separated by a minimum distance $d = 340\:$nm, for combined polarities $p=p_1p_2=+1$ and $p=-1$.
The eigenfrequencies of these disks are $\omega_0/2\pi = 0.56\:$GHz for $R = 150\:$nm  and $\omega_0/2\pi = 0.49\:$GHz for $R = 170\:$nm and were obtained using the two-vortex model \cite{Guslienko:2001, Metlov:2002}. These values are in good agreement with those obtained from micromagnetic simulation, respectively, $\omega_0/2\pi = 0.58\:$GHz and $\omega_0/2\pi = 0.52\:$GHz.
The eigenfrequencies of the interacting
pair as a function of distance are shown in Fig. \ref{fig:CouplingVsD}. It is apparent that for increasing $d$ the frequencies
tend to the values $\omega_i$ of the isolated disks. Note also how the frequencies are dependent on the relative polarities of the disks. These results are in a quantitative agreement with micromagnetic simulations of the eigenfrequencies, showing that this method
is consistent; for example, for $d=340\,$nm from Eq. \ref{eq:omega} we find $\omega^1_+/2\pi=0.57\,$GHz, $\omega^1_-/2\pi=0.48\,$GHz and
from the simulation $\omega^1_+/2\pi=0.57\,$GHz, $\omega^1_-/2\pi=0.51\,$GHz.

\subsection{Thiele's equation and the magnetic vortex echo}
\label{subsection:ThieleEcho}

An alternative way of studying the time dependent magnetization and interaction between two magnetic nanodisks with different diameters is to use a new phenomenon, the magnetic vortex echo (MVE), described recently \citep{Garcia:2012}. MVE is an effect of the vortex gyrotropic motion around an equilibrium position, and arises from the refocusing of the overall magnetization of the ensemble containing many nanoelements. MVE can be used as a tool to characterize nanostructures that exhibit a vortex ground state as regards the homogeneity and intensity of the interaction between its elements, properties that are relevant for device applications, as explained by Garcia \textit{et al.} \cite{Garcia:2012}. In the previous work, the system studied was a matrix of nanodisks, here we have applied the
method to pairs of nanodisks.

In order to observe the MVE one needs an ensemble of nanoelements with a distribution of gyrotropic frequencies (or distribution of
diameters). In our case we used an ensemble of 50 pairs of Permalloy nanodisks of different diameters, with constant center to center distance. We used disks with $20\,$nm thickness, and an approximately Gaussian distribution of diameters (average diameter of $D=250$ nm, and $\sigma=10$ nm); the 50 pairs were formed with disks of the same ensemble chosen in Garcia
 {\textit et al.}\cite{Garcia:2012}.

\begin{figure}[!ht] \centering \includegraphics[width=1\columnwidth]{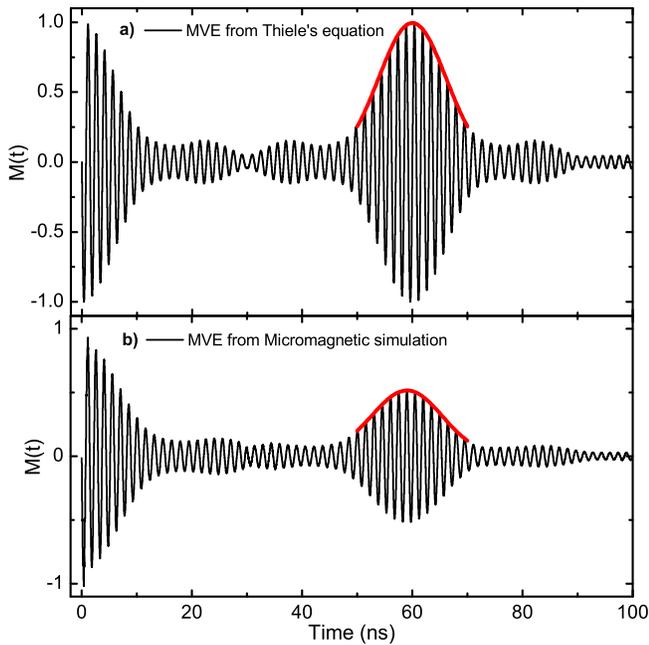} \caption{Magnetization $M(t)$ of an ensemble of disk pairs separated by 450$\:$nm, versus time, showing the initial decay and the refocalization of the rotating magnetic cores at $t=60\:$ns - the magnetic vortex echo (MVE). The red lines show the computer fits to the echo, used to derive the values of $T^*_2$. In a) $M(t)$ was obtained by solving Thiele's equation, and in b) $M(t)$ was obtained
from a micromagnetic simulation using the OOMMF code.}\label{fig:EchoPairs1} \end{figure}

To obtain an echo, an external magnetic field, with 25\;mT intensity in the $y$ direction, was applied to each pair of disks, displacing the vortex cores in the $x$ direction; withdrawing the field, the cores start to precess, performing a gyrotropic motion. The defocusing of the motion of the disks, due to the distribution of diameters (consequently, distribution of frequencies of width $\Delta\omega$) leads to a decay of the total magnetization $M(t)$. After a time $\tau$, a magnetic pulse, with 300\;mT intensity in the $z$ direction and duration of 100\;ps, inverts the polarity of the disks; after the pulse, the refocusing produces the MVE, as shown in Fig. \ref{fig:EchoPairs1}. The decay of the total initial magnetization due to this defocusing, as well as the decay of the echoes, are characterized by a relaxation time $T_2^*$, which depends on the standard deviation $\Delta\omega$, on the Gilbert damping constant $\alpha$ and on the interaction between the neighbor disks as\cite{Garcia:2012}:

\begin{equation} \frac{1}{T_2^{*}}=\Delta \omega + \frac{1}{T_2}= \Delta \omega + \frac{1}{T_2^{ \prime}} + \frac{1}{2T_{\alpha}} \label{eq:T2*} \end{equation}
 \noindent where $2T_{\alpha}$ is the relaxation time related to the damping constant $\alpha$, and $1/T_2^{ \prime}$ accounts for the interaction between the disks.

Therefore, for an ensemble of disks with the same $\alpha$, $1/T_2^{*}$ varies linearly with the strength of the interaction between them.

To obtain the vortex echo we have used Thiele`s equation to compute the coupling frequencies of the pairs of magnetic nanodisks with different diameters. The individual eigenfrequencies were computed within the two-vortex model, which is known to give more accurate results \cite{Guslienko:2002}. For each separation between the two disks we computed the variation of the magnetization as a function of time; the contributions of all the 50 pairs of disks were then added. The result, after the application of the external pulse and the formation of the echo,
is shown in Fig. \ref{fig:EchoPairs1}a.

\begin{figure}[!ht] \centering \includegraphics[width=1\columnwidth]{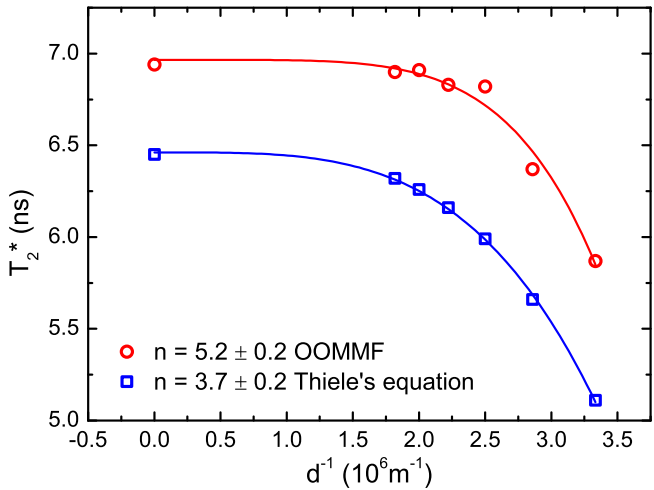} \caption{Graph of $T_2^{*}$ obtained from the fitting of the magnetic vortex echoes for pairs of disks with a gaussian distribution of diameters. The continuous lines are computer fits to a function $a+b*d^{-n}$; a) red circles represent
values of $T_2^{*}$ derived, for each distance, from the echoes obtained from micromagnetic simulation (fit with $n=5.2\pm 0.2)$; b) blue squares represent $T_2^{*}$ from the echoes
generated using Thiele's equation ($n=3.7\pm 0.2$).}\label{fig:T2starPairs} \end{figure}

The relaxation times
$T_2^{*}$, that also measure the interaction strength, derived using two methods, based on the magnetic vortex echoes, are comparable, differing by about $10\%$ (Fig. \ref{fig:T2starPairs}).

\subsection{Micromagnetic simulation and the magnetic vortex echo}
\label{section:simulation}

The micromagnetic simulations were made for
 Permalloy nanodisks, cells of $5 \times 5 \times 20\,$nm$^3$, $\alpha=0$, using the OOMMF code \cite{Donahue:1999}. Each pair of disks was simulated individually and the overall magnetization of the 50 pairs were obtained by simply summing up the contributions of all the pairs. The echoes of the system, in which the distance between the disks was varied in the range 260-550$\,$nm, were then simulated.

The vortex echo obtained from the micromagnetic simulation is illustrated in Fig. \ref{fig:EchoPairs1}b; note that the echo intensity in this simulation is smaller than
the magnetization at $t=0\,$ns, whereas the echo generated from Thiele's equation (Fig. \ref{fig:EchoPairs1}a) shows no reduction in the echo intensity. This difference is associated to the energy that
is transferred to the spin waves during the process of inversion of polarity in the micromagnetic simulation.

The curves of $T_2^{*}$ versus disk separation obtained from the magnetic vortex echoes and
 the two methods - Thiele's equation and micromagnetic simulation (Fig. \ref{fig:T2starPairs});  the results corresponding to infinite separation were computed applying the method to individual disks.

The curves obtained with the two different methods (Fig. \ref{fig:T2starPairs}) show the same qualitative behavior, and a quantitative difference of about $10\%$. The vortex echo analytical results are sensitive to the input eingenfrequecies.
 The difference in the values of $n$ (of about $30\%$) is apparently due to the fact that the higher order terms of the multipolar expansion of the magnetic field are more relevant in the micromagnetic simulation.
 Another source of difference between the two results may arise from the fact that the rigid-vortex model only takes into account the surface charges, whereas on the micromagnetic simulation the volume charge are also computed.

\section{Conclusions}

In this paper we have studied the interaction between pairs of magnetic nanodisks of different diameters and vortex ground state; from an ensemble of
magnetic nanodisks with a gaussian distribution of diameters, we created fifty pairs of nanodisks. In this study we have a)
derived analytically the expressions of the coupling integrals $I_x$ and $I_y$ that describe this interaction; b) from the time dependent magnetizations
derived from the numerical solution of Thiele's equation we applied the
vortex echo method\cite{Garcia:2012} to derive the dependence of the interaction with distance; c) we made a micromagnetic simulation to obtain $M(t)$ and
again applied the echo method to evaluate the strength of the interaction between the disks.
We have also obtained the variation with distance between the disks, of the coupling frequencies, derived from Thiele's equation.

The coupling integrals $I_x$ and $I_y$ vary depending on distance in a way comparable to the results obtained by other authors. The relaxation times
$T_2^{*}$, that also measure the interaction strength, derived using two methods based on the magnetic vortex echoes, are comparable, differing by about $10\%$ (Fig. \ref{fig:T2starPairs}).
The fitting to the $T_2^{*}$ curves obtained from these two techniques show an approximate dependence of the form $\propto d^{-n}$, with values of $n$ that vary between $5.2\pm 0.2$ (micromagnetic simulation) and $3.7\pm 0.2$ (Thiele's equation),
comparable to other results of coupling between magnetic vortex disks in the literature.

\begin{acknowledgments}

The authors would like to thank the support of the Brazilian agencies CNPq, FAPERJ.

\end{acknowledgments}

\vspace{10mm}

%\bibliographystyle{unsrt}
%\bibliography{Nanomagnetism}

\end{document}